# Image transmission over OFDM channel with rate allocation scheme and minimum peak-to-average power ratio

Usama S. Mohammed, H. A. Hamada

**Abstract**— This paper proposes new scheme for efficient rate allocation in conjunction with reducing peak-to-average power ratio (PAPR) in orthogonal frequency-division multiplexing (OFDM) systems. Modification of the set partitioning in hierarchical trees (SPIHT) image coder is proposed to generate four different groups of bit-stream relative to its significances. The significant bits, the sign bits, the set bits and the refinement bits are transmitted in four different groups. The proposed method for reducing the PAPR utilizes twice the unequal error protection (UEP), using the Read-Solomon codes (RS), in conjunction with bit-rate allocation and selective interleaving to provide minimum PAPR. The output bit-stream from the source code (SPIHT) will be started by the most significant types of bits (first group of bits). The optimal unequal error protection (UEP) of the four groups is proposed based on the channel destortion. The proposed structure provides significant improvement in bit error rate (BER) performance. Performed computer simulations have shown that the proposed scheme outperform the performance of most of the recent PAPR reduction techniques in most cases. Moreover, the simulation results indicate that the proposed scheme provides significantly better PSNR performance in comparison to well-known robust coding schemes.

——————————— ◆ ———————————

## 1 INTRODUCTION

Orthogonal frequency-division multiplexing (OFDM) scheme [1], [2] is used in most recent wireless communications systems due to its high spectrum efficiency and robustness in multi-path propagation. OFDM is a special form of multi-carrier modulation and mitigate inter symbol interference (ISI) by multiplexing the data on orthogonal property. Moreover, it is, spectrally, more sufficient technique than a conventional signal carrier modulation technique. However, one of major drawbacks of OFDM is the high peak-to-average power ratio (PAPR) of the transmitted signal [3]. Several methods have been proposed to reduce the PAPR. These methods can be categorized into signal distortion methods and signal scrambling methods [4-7]. The signal distortion methods reduce high peaks directly by distorting the signal prior to amplification. Clipping the signal before amplification is a simple method to limit PAPR. However, the out-of-band- and in-band interference due to use these methods will increase the degradation of the system performance. Signal scrambling methods are all variations on how to scramble the codes to decrease the PAPR. There are three practical solutions of the signal scrambling methods [4]: block coding, selective mapping and partial transmit sequences. The signal scrambling techniques can be classified into schemes with explicit side information and schemes without side information. The scheme of signal scrambling with side information introduces redundancy, so the effective throughput is reduced. At this point it is worth mentioning that increasing redundancy affects on the total transmission rate. So, the maximum redundancy for each data packet must be estimated relative to the packet data significance. As an illustrative method, the joint source and block coding scheme with bit-rate optimization. In this paper, the proposed method for reducing the PAPR utilizes twice the unequal error protection (UEP) using the Read-Solomon codes (RS) [14] in conjunction using selective interleaving with bit-rate allocation. In OFDM system, a block of N complex symbols is formed with each symbo modulation one of N subcarriers with frequency for  The N-subcarrier is chosen as orthogonal. The complex envelope of the transmitted OFDM signal is represented by [2]:

$$x(t) = \frac{1}{N} \sum_{n=0}^{N-1} S_n e^{j2\pi f_n t} \qquad (1)$$

Where $S_n$ is the data symbol, N is number of subcarriers and $f_n$ represents frequency of $n$-th subcarriers, the OFDM system shown in Fig. 1.

In this work, a new scheme combines simply modification of the set partitioning in hierarchical trees (SPIHT) image coding technique followed by an optimum UEP technique is proposed.

————————————————

- *Usama S. Mohammed Auther is with the Electrical Engineering Department, Assiut University, Assiut 71516, Egypt.*
- *H. A. Hamada Author is with the Electrical Engineering Department, South Valley University, Aswan 81542, Egypt.*

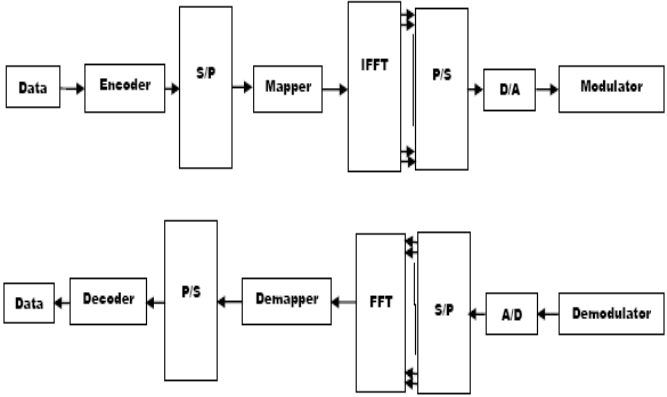

Fig. 1. Block description of the OFDM system

The modified SPIHT coder will generate four groups of bit-stream. The significant bits, the sign bits, the set bits, and the refinement bits are transmitted in four different groups. The main idea behind this approach is that the output of the SPIHT image coding (source code) will be sent relative to its significant information. The idea of the proposed algorithm can be used with any scalable image coding technique. The paper is organized as follows: the source code and the PAPR problem in OFDM are discussed in the next section. The proposed algorithm is introduced in Section 3. Selective interleaving is presented in section 4. Implementation and computer results are presented in section 5. Finally, conclusions are given in Section 6.

## 2 SOURCE CODE AND PEAK-TO-AVERAGE POWER RATIO PROBLEM

### 2.1 Peak-to-Average Power Ratio Problem

OFDM consist of many modulated subcarriers. As mentioned in the previous section, this leads to a problem with the peak to average power ratio. If N subcarriers are added up coherently, the peak power is N times the average power in the case of the baseband signal. The PAPR is defined [7], [8] as follows:

$$\text{PAPR} = \frac{\max\left\{|x(t)|^2\right\}}{E\left\{|x(t)|^2\right\}} \quad (2)$$

It is clear, from Equation (2), that the PAPR reduction techniques are concerned with reducing max. However, since most systems employ discrete-time signals, the amplitude of samples of x(t) is dealt with in many of the PAPR reduction techniques. The various approaches are quite different from each other and impose different constraints. First, coding in [9] relies on using several bits or bit sequences which would carry a properly chosen code (sometimes with error correcting capabilities) that minimizes the PAPR of the resulting transmitted signal. The PAPR is reduced, but so is the data rate. Other methods use phase manipulations (e.g. Selective Mapping (SLM), Partial Transmit Sequences (PTS), Random Phasor [11-13]).

To reduce PAPR based on the properties of the μ-law compander to decrease dynamic range of the signal [23] is used. The compander consists of compressor and expander. The compressor is a simple logarithm computation. The reverse computation of a compressor is called an expander. In this work, the compression at the transmit end after the IFFT process and expansion at the receiver end prior to FFT process are used. There are two types of companders that are used here which are described in details in [21]. These two types are μ-law and A-law companders.

### 2.2 Source Coder (SPIHT) and the UEP Process

The SPIHT method is consider the best image coding technique in terms of decoded image quality, progressive rate control and transmission, and the simplicity of the coding process [13]. In the SPIHT coding algorithm, after the wavelet transform using 9/7 tap wavelets from Antonini et al. [15] is applied to an image, the main algorithm works by partitioning the wavelet decomposed image into significant and insignificant partitions.

In this work, modification of the output bit-stream of the SPIHT coder is done. The modification process is based on the type of bits and their contribution in the PSNR of the reconstructed image. The bit error sensitivity (BES) study is performed by first coding the original image using the SPIHT coder. One bit in the coded image is corrupted, starting from the first bit to the last bit. Each time a bit is corrupted, the coded image is decoded and the resultant MSE is obtained. The corrupted bit is corrected before proceeding on to the next bit. On analysis, there are 4 major types of bit sensitivities within the SPIHT coded bits. Their description is summarized as follows: (1) the significance bit in the bit stream. It decides whether nodes in the LIP are significant, (2) the sign bit of a significant node that is transmitted after the significance bit, (3) the set bit that decided the set is significant or not, (4) the refinement bits that are transmitted during the refinement passes.

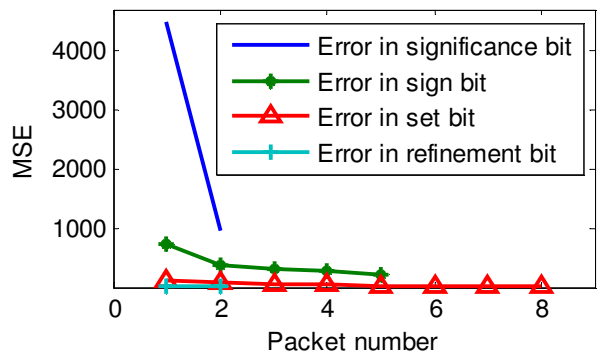

Fig. 2. Error bit sensitivities within the SPIHT coded bit stream

Fig.2, show the order of significance from the most significant types of bits to the least significant for gray (256*256) Lena image is: significance bits > sign bits > set bits > refinement bits. In the first step of the proposed scheme, SPIHT coder will be modified to generate four

groups of bit stream related to the order significance i.e.; the output bit stream will be started by the most significant types of bit (first group of bits).

The optimum bit rate for each group of bits is computed using the optimization algorithm. The inputs to the optimization algorithm are the packet length, the bit error rate (BER), and the expected decrease in the distortion of packets. The output rate allocation vectors are used with RS codes to generate the transmitted bit-stream.

## 3 PROPOSED OPTIMUM DOUBLE CODING TECHNIQUE FOR OVER ALL DISTORTION REDUCTION

OFDM system use RS (Reed-Solomon) coder to protect data transmits. In this work, we investigate two RS block codes to construct product block codes. The product codes have been introduced in [12]. They permit the construction of long correcting code with high correction capability using two elementary codes. The product coding P=C1*C2 is obtained by placing k1k2 information bits in an array of k1 rows and k2 columns. A first coding is operated (by C1 code) on the k2 columns each one with the length k1 which lead to the matrix P',(Figure 1). A second coding (by C2 code) is done on the n1 rows of matrix P' (with length n1k2) to give the final matrix P (with length n1n2).

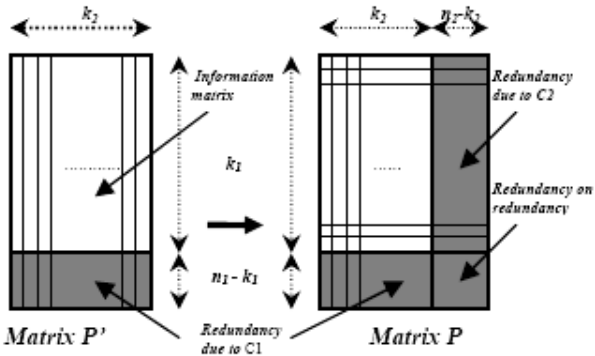

Fig. 3. Product coding schem

A comprehensive review of the great variety of error control and concealment techniques has been presented in the paper by Wang and Zhu [16]. Unequal error protection (UEP) [17] can significantly increase the robustness of the transmission and provide graceful degradation of the picture quality in case of a deteriorating channel. UEP was adapted to packet networks by Albanese et al. in their priority encoding transmission (PET) scheme. Optimization of the rate allocation in such a scheme was addressed by [18]-[19].

In this work, the problem of UEP will be studied from different view. The source code (SPIHT for image) is modified to generate four different bit-streams. The contribution of each one in the overall peak signals – to – noise ratio PSNR in the received image is estimated and the protection of each one is based on this estimation.

### 3.1 Problem Formulation

In this work, the problem of FEC using RS codes will be introduced. We assume that the image source is encoded by the SPIHT coder. The generated bitstream is partitioned into a sequence of packets.

Let $\Delta D_i \geq 0$ denote the expected decrease in distortion if the $i^{th}$ packet is decoded. The overall distortion can be written as follows:

$$D(l) = D_0 - \sum_{i=1}^{l} P_i \Delta D_i \qquad (3)$$

Where $D_0$ is the expected distortion when the rate is zero, $P_i$ is the probability that the $i^{th}$ packet and its preceding packets are received correctly, and $l$ is the number of source packets that the transmitter chose to send. The probability $P_i$ can be written in the form:

$$P_i = \prod_{j=1}^{i} Q_j(r_j) \qquad (4)$$

Where $Q_j(r_j)$ is the probability that the $j^{th}$ layer of source packet is received correctly when sending by a rate of $r_j$. Substituting from equation (4) in equation (3) yields to:

$$D(r) = D(l) = D_0 - \sum_{i=1}^{l} (\prod_{j=1}^{i} Q_j(r_j)) \Delta D_i \qquad (5)$$

With the distortion expression in equation (5), for any rate allocation vector $r_j$, we can minimize the expected distortion subject to a transmission rate constraint. The problem can be formulated as follows:

$$\min_r D(r) \quad \text{subject to} \quad \sum_{j=1}^{l} r_j \leq R \qquad (6)$$

Where R is the total transmission rate.

### 3.2 Forward Error Correction with RS Codes

Assuming that, the stream is partitioned into coding blocks. Each coding block has k source packet. For each block of k source packets, we assume that $N-K$ parity packets are produced using a systematic $(N,K)\ RS$ style f redundancy that will needed by the transmitter to protect the source. . In this work, we study the signal transmission over two type of erasure correction code. $(N-K)$ is the maximum amount of channel, BSC and AWGN channel. First, let the packet transmitted through binary symmetric channel with bit error rate . Then, the packet loss probability is given by:

$$S(P) = 1 - [1-P]^m \qquad (7)$$

Assuming independent bits and $m$ is the number of bits in the packet. For AWGN channel packet is transmitted symbol by symbol through the channel, where each MQAM symbol has $b$ bits in it, modulated using fixed power MQAM. Thus, each packet corresponds to $L/b = L_s$ MQAM symbols. We assume additive white Gaussian noise $(AWGN)$ at the receiver frontend, and no interference from other signals. The channel is narrowband, so the power spectra of both the received signal and the noise have no frequency dependence, i.e., the channel is characterized by a single path gain variable.

For $AWGN$

$$S(b, \gamma_s, L) = [1 - P_e(b, \gamma_s)]^{L \cdot f_h} \quad (8)$$

Where $\gamma_s$ is the $SNR$ per symbol, $P_e$ is the symbol error rate where $P_e$ of MQAM in $AWGN$ channels is (approximately) given by [20].

$$P_e(b, \gamma_s) = 4(1 - 2^{-b/2}) Q\left(\sqrt{\frac{3}{2^b - 1} \gamma_s}\right) \quad (9)$$

It is shown in [22] that the output of a linear minimum mean-square error ($MMSE$) detector is approximated by a Gaussian distribution. After channel decoding with a $(N, K)$ RS code for source layer $j$, the probability that the $j^{th}$ layer of source packet is received correctly can be written as follows [21]:

$$Q_j(r_j) = \frac{EP(r_j, k, S_j(P))}{k} \quad (10)$$

Where $EP(r_j, k, S_j(P))$ is the expected number of source packets that can be recovered and it can be written as follows:

$$EP(r_j, k, S_j(P)) = \sum_{v=1}^{k-1} \binom{r_j}{v} S_j(P)^{r_j - v} (1 - S_j(P))^v \left(\frac{v}{r_j}\right)$$
$$+ \sum_{v=k}^{r_j} \binom{r_j}{v} S_j(P)^{r_j - v} (1 - S_j(P))^v k \quad (11)$$

Hence, with the expected distortion expression in equations (5), (9) and (10), for any rate allocation vector $r_j$, we can optimize the rate vector to minimize the expected distortion subject to a transmission rate constraint.

## 3.3 The Optimization Technique.

Equation (6) can be solved by finding the rate allocation vector r that minimizes the Lagrangian equation.

$$J(r, \lambda) = D(r) + \lambda \sum_{i=1}^{l} r_i \quad (12)$$

OR

$$J(r, \lambda) = D_0 + \sum_{i=1}^{l} [(-\prod_{j=1}^{i} Q_j(r_j)) \Delta D_i + \lambda r_i] \quad (13)$$

The solution of this problem is characterized by the set of distortion increment ⃞Di and Qj(r) with which the jth layer source packet is recovered correctly. In this work, the problem is solved by using an iterative approach that is based on the method of alternating variables [25]. The objective function J(r1,……,rl) in equation (12) is minimized one variable at a time, keeping the other variables constant, until convergence. To be specific, let r(0) be any initial rate allocation vector and let r(t) =(r1(t),……, rl(t)) be determined for t=1,2,… as follows: select one component x ε {r1,……,rl} to optimize at step t. this can be done in a round-robin style. Then, for x = ri we can perform the following rate optimization:

$$r_i^{(t)} = \arg\min_{ri} J(r_1^{(t)}, ......, r_l^{(t)})$$
$$= \arg\min_{ri} \sum_{v=i}^{l} (-\prod_{j=1}^{v} Q_j(r_j)) \Delta D_v + \lambda r_i \quad (14)$$

For fixed ⃞ the minimization problem can be solved using standard non-linear optimization procedures, such as gradient-descent type algorithm [24]. In our simulation results, we always start with the initial rate allocation vector r= (1, 1,……,1).

## 3.4 The Proposed Scheme

In the beginning as shown in Fig. 4, the SPIHT coder output bitstream modification has done. The modification process is based on the type of bits and their contribution in the PSNR of the reconstructed image. The bit error sensitivity (BES) study is performed by first coding the original image using the SPIHT coder. One bit in the coded image is then corrupted, starting from the first bit to the last bit. Each time a bit is corrupted, the coded image is decoded and the resultant MSE is calculated. The corrupted bit is corrected before proceeding on to the next bit. On analysis, the resultant BES study is carried out on a *256\*256\*8* image of Lena coded at source code rate of 0.5 bpp using the SPIHT algorithm. On analysis, there are 4 major types of bit sensitivities within the SPIHT coded bits, as shown in Fig.2.

The output bitstream will be started by the most significant types of bits (first group of bits).

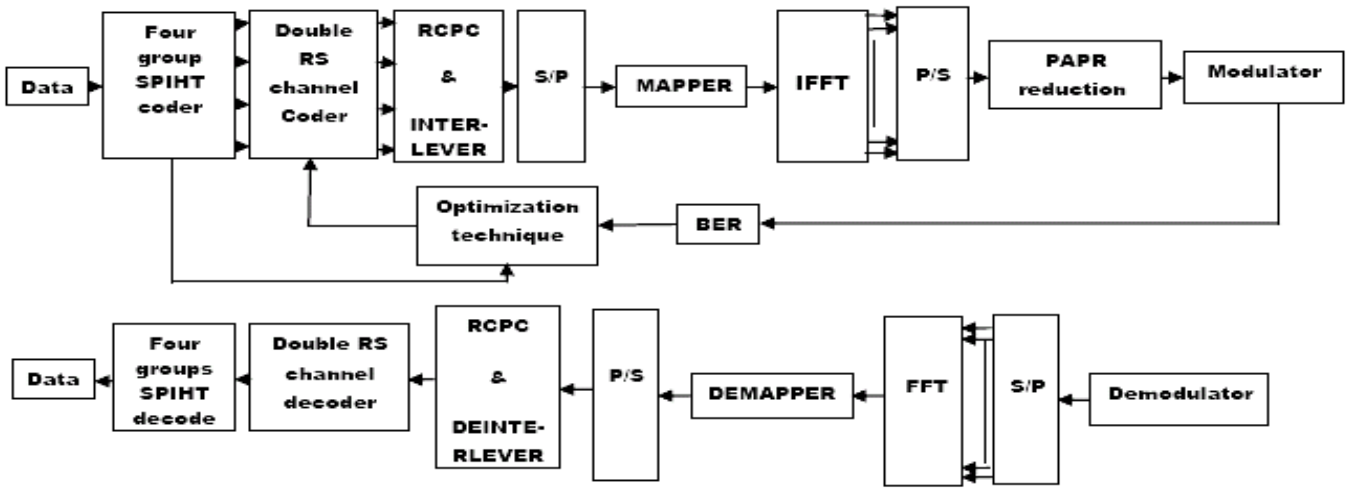

Fig.4. Modified OFDM system

The proposed scheme can be summarized as follows: *(1)* the SPIHT image coding technique is modified to generate 4 groups of bitstream related to the order of significance, *(2)* the output data is partitioned into a sequence of packets, *(3)* the changing in MSE is calculated and the expected decrease in distortion $\Delta D_i$ is then approximately estimated, *(4)* the optimization algorithm is applied to generate the optimum bit rate for each group of bits. The inputs to the optimization algorithm are the packet length, the bit error rate ($BER$), and expected distortion $\Delta D_i$, *(3)* the output are the rate allocation vectors used with RS codes to generate the transmitted bitstream, *(4)* the receiver will decode the receiving bitstream by using RS decoder and the modified SPIHT decoder.

### 3.5 IMAGE TRANSMISSION OVER OFDM SYSTEM WITH RS CHANNEL CODING

The proposed scheme is tested on *BSC* and *AWGN* channel and each experiment is repeated 50 times at a given channel $BER$ of 0.001, for $256 \times 256$ gray-scale Lena image. The modified SPIHT coder applied to the wavelet coefficients to generate the source bitstream with a bit rate of 0.5 bpp. The source bitstream is divided into packets of length 2000 bits and each packet is divided into 25 blocks of length 80 bits. This means that each block has 10 symbols of one byte for each. In the simulation results, the total bitstream of the Lena image is divided into 17 packets. These 17 packets are divided into 4 groups of packets as follows: (1) 2 packets for the significant information, (2) 5 packets for the sign bit, (3) 8 packets for the set bit, (4) 2 packets for the refinement bits. In the beginning, we tried to put the results of the $UEP$ method in a comparison form with the result of the equal error protection ($EEP$) method. In the case of $EEP$, the number of RS symbols is selected to be 8-symbol for each row of packet that make the total symbols per packet are 450. In the case of $UEP$, the total symbols per packet ($R$ in Eq.7) is selected to be no more than 450 symbol per packet and the optimization algorithm is used to determine the number of RS symbols for each column and row of packets.

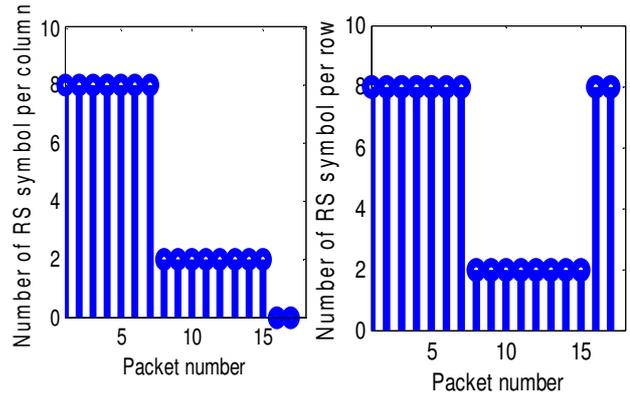

Fig. 5 Channel rates for the protection of "Lena" as determined using the UEP algorithm

The result of this step is shown in Fig.5. It is clear from Fig. 5 that the significance and sign packets are protected by 8-symbol for column and row, the set packets are protected by 2-symbol for column and row, and the refinement packets are protected by 8-symbol for row and zero for column.

Fig. 6 depicts the average PSNR of the decoded Lena image as a function of the transmission rate for the EEP and the UEP coding method. In particular, our proposed scheme for *UEP* outperforms the system of EEP by 4.2 dB.

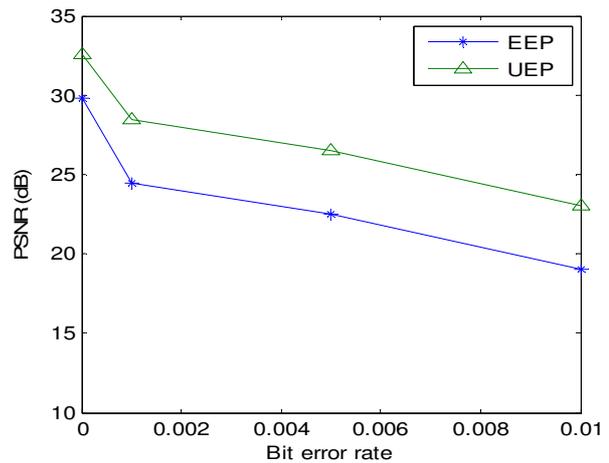

Fig.6. the average PSNR of the decoded LENA image as a function of the channel BER for EEP & UEP channel coding.

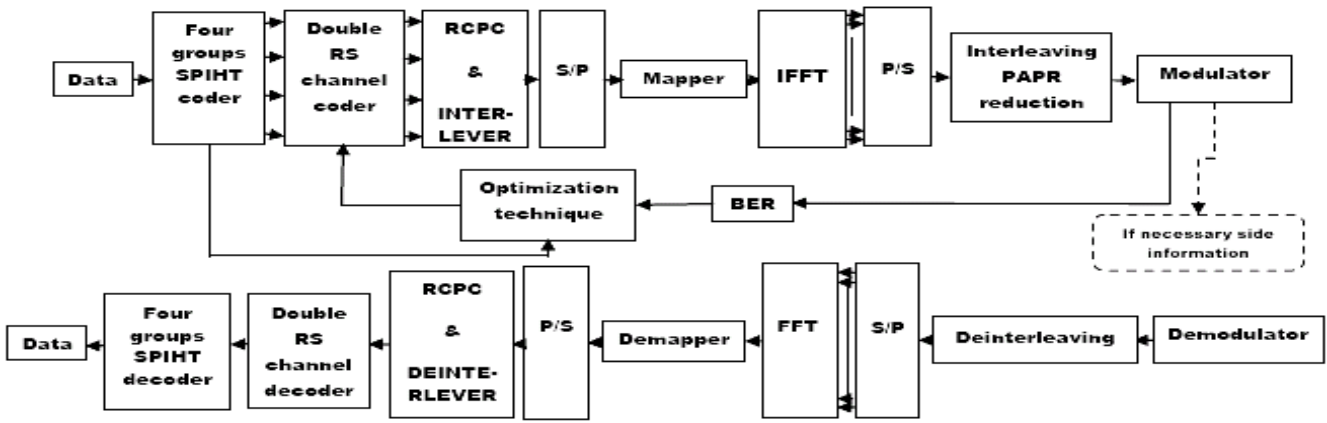

Fig. 7. Modified OFDM use interleaving selection

## 4. INTERLEAVER SELECTOR

The detailed descriptions of interleaver are found in [21]. In The proposed approach, k interleave are used at the transmitter. This interleaver produces K permuted frames of the input data sequence. These permutations can be done either before or after the modulation (mapping). The minimum PAPR frame of all the K frames is selected for transmission. The identity of the corresponding interleaver is also sent to the receiver as side information. Fig. 7 show the interleaving selection method the data will be interleaving and then measure the PAPR and if it is large than the PAPRo the data will bass throw multi type of interleaving and select the interleaving block which gave minimum PAPR, The interleaving PAPR reduction block flow chart in Fig. 8.

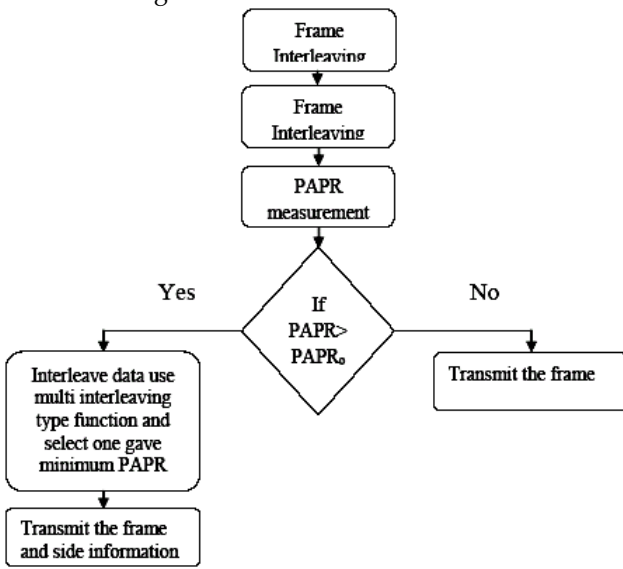

Fig.8 Selective interleaving PAPR reduction block flow chart

## 5 COMPUTER RESULTS

We compare between the data result of transmit signal over original and modified OFDM systems using Compound method for PAPR reduction method and PAPR reduction using selective interleaving.

### 5.1 compound method simulation result.

In this work we use the compound method µ=2 for original and modified OFDM system data will pass over two type of channel BSC and AWGN channel.

### A) BS Channel:

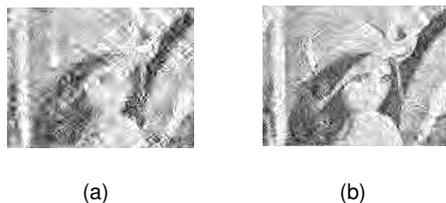

(a)          (b)

Fig.9 the receved Lena image transmit over original and modified OFDM system over BSC with BER=0.001 (PSNR result is proposed in dB), (a) original OFDM (PSNR= 17.973), (b) modified OFDM (PSNR= 20.7454)

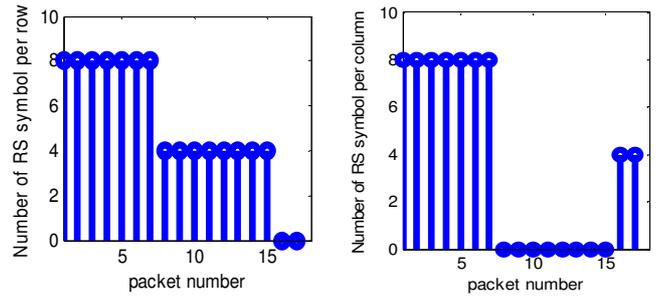

Fig. 10 Channel rates protection as determined using the $UEP$ algorithm for Lena image over BSC with BER=0.01.

TABLE 1
SIMULATION RESULT WHEN TRANSMITTED DATA OVER OFDM SYSTEMS USEING COMPOUND METHOD WITH EEP AND UEP, OVER BSC CHANNEL.

| system | MSE | transmission rate | Max (PAPR) | frame No. | $PAPR_{av}$ (dB) |
|---|---|---|---|---|---|
| Channel BER=0.01 | | | | | |
| Original OFDM | 1.1667 e+004 | 122880 bps | 8.5378 dB | 96 | 6.28 |
| Modified OFDM | 4.0651 e+003 | 121600 bps | 7.5253 dB | 95 | 6.05 |
| Channel BER=0.001 | | | | | |
| Original OFDM | 1.0369 e+003 | 122880 bps | 8.5378 dB | 96 | 6.277 |
| Modified OFDM | 5.4769 e+002 | 117760 bps | 8.6578 dB | 92 | 6.05 |

### B) AWGN channel:

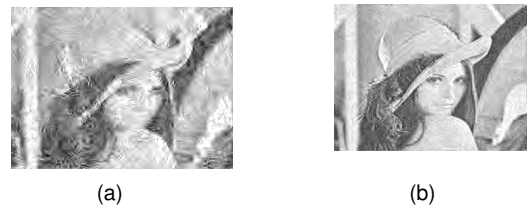

(a)          (b)

Fig.11 the received Lena image transmit over original and modified OFDM system over AWGN channel with BER=0.001 (PSNR result is proposed in dB), (a) original OFDM (PSNR= 21.6948), (b) modified OFDM (PSNR= 26.13)

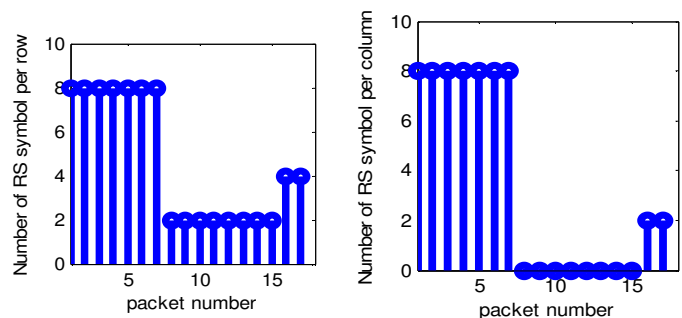

Fig.12 Channel rates protection as determined using the UEP algorithm for Lena image over AWGN channel with BER=0.001.

TABLE 2
SIMULATION RESULT WHEN DATA TRANSMITTED THROWS SYSTEMS WITH EEP AND UEP WITH COMPOUND METHOD OVER AWGN CHANNEL.

| system | MSE | transmission rate | Max (PAPR) | Frame No. | PAPR$_{av}$ (dB) |
|---|---|---|---|---|---|
| Channel BER=0.01 | | | | | |
| Original OFDM | 4.3469e+003 | 30736 syp/s | 7.6733 dB | 96 | 6.27 |
| Modified OFDM | 1.55262e+003 | 29455 syp/s | 7.6733 dB | 92 | 6.16 |
| Channel BER=0.001 | | | | | |
| Original OFDM | 4.4014+002 | 30736 syp/s | 8.5378 dB | 96 | 6.28 |
| Modified OFDM | 1.585e002 | 28480 syp/s | 7.7989 dB | 89 | 6.14 |

## 5.2 SELECTIVE INTERLEAVING

Here we use the compound method µ=2 for original and modified OFDM system data will pass over two type of channel BSC and AWGN channel.

### A) BS Channel:

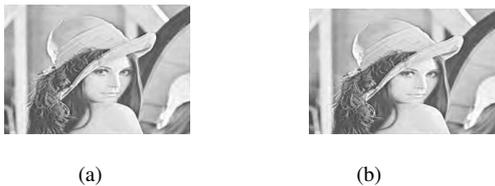

(a)        (b)

Fig. 13 the receved Lena image transmit over original and modified OFDM system used interleaving selection with channel BER=0.001 (PSNR result is proposed in dB), (a) original OFDM (PSNR= 32.542), (b) modified OFDM (PSNR= 32.542)

TABLE 3
SIMULATION RESULT WHEN TRANSMITTED DATA OVER OFDM SYSTEMS USEING SELECTIVE INTERLEAVING with EEP AND UEP, OVER BSC CHANNEL.

| System | MSE | transmission rate | Max(PAPR) | Frame No. | PAPR$_{av}$ (dB) |
|---|---|---|---|---|---|
| Channel BER = 0.01 | | | | | |
| Original OFDM | 1.27e+003 | 123196 bps | 7.6828 dB | 96 | 6.73 |
| Modified OFDM | 5.12e+002 | 119356 bps | 7.5267 dB | 93 | 6.72 |
| Channel BER = 0.001 | | | | | |
| Original OFDM | 32.5 | 123196 bps | 7.6828 dB | 96 | 4.73 |
| Modified OFDM | 32.5 | 120624 bps | 7.5267 dB | 94 | 4.71 |

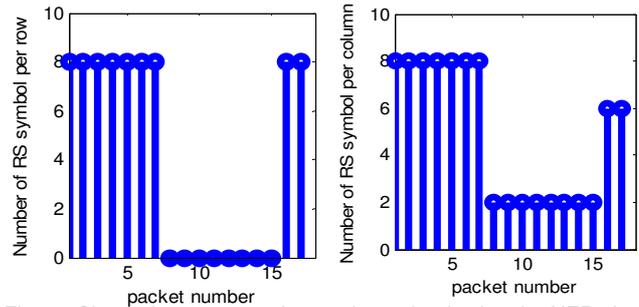

Fig. 14 Channel rates protection as determined using the UEP algorithm

### B) AWGN Channel:

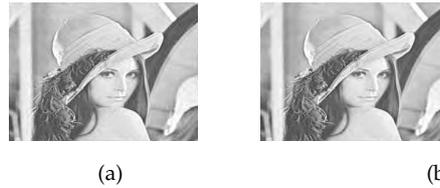

(a)        (b)

Fig. 15 the receved Lena image transmit over original and modified OFDM system used interleavig selection with channel BER=0.001 (PSNR result is proposed in dB), (a) original OFDM (PSNR= 32.02), (b) modified OFDM (PSNR= 32.542)

TABLE 4
SIMULATION RESULT WHEN TRANSMITTED DATA OVER OFDM SYSTEMS USEING SELECTIVE INTERLEAVING with EEP AND UEP, OVER AWGN CHANNEL.

| System | MSE | transmission rate | Max (PAPR) | Frame No. | PAPR$_{av}$ (dB) |
|---|---|---|---|---|---|
| Channel BER = 0.01 | | | | | |
| Original OFDM | 6.221e+002 | 30799 syp/s | 7.6828 dB | 96 | 6.734 |
| Modified OFDM | 3.146e+002 | 30480 syp/s | 7.5267 dB | 95 | 6.71 |
| Channel BER = 0.001 | | | | | |
| Original OFDM | 40.8309 | 30799 syp/s | 7.6828 dB | 96 | 6.734 |
| Modified OFDM | 36.2212 | 28552 syp/s | 7.8314 dB | 89 | 4.714 |

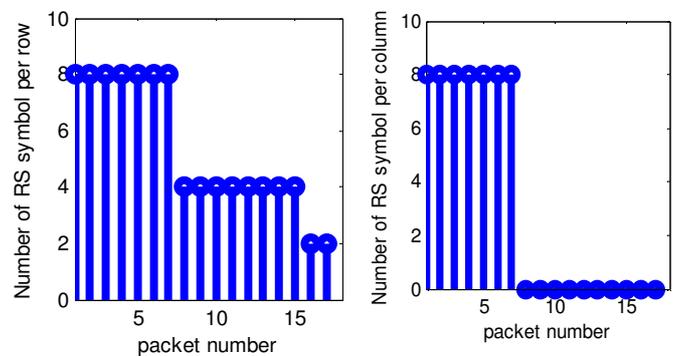

Fig. 16 Channel rates protection as determined using the UEP algorithm

Fig. 16 show the power Complementary Cumulative Distribution Function (CCDF) curves provide critical information about the signal encountered in Interleaving selection and the another PAPR reduction method as selective mapping (SLM = 4)[11], Barlett weighting function [26], Clipping method, Huffmant coder [5], Compound function [23], and Selective Interleaving.

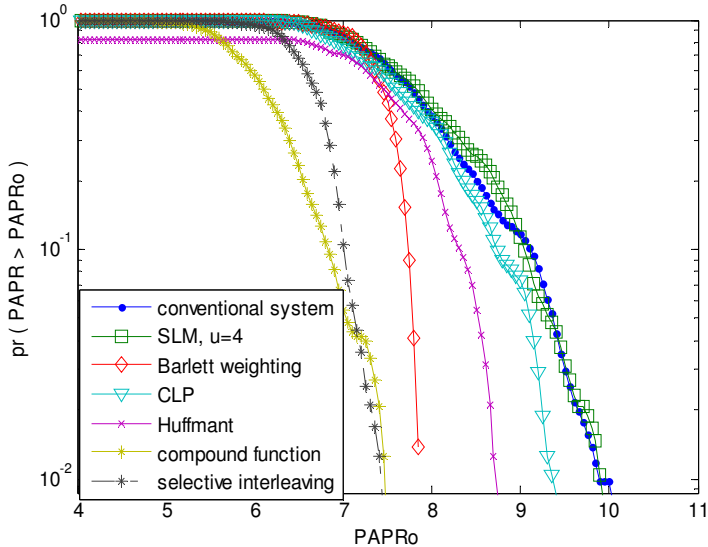

Fig. 17 CCDF Comparison of the Interleaving selection method with PAPR reduction method

## 6 CONCLUSION

In this paper we have consider the problem of the PAPR reduction in OFDM system in conjunction with the efficient rate allocation scheme. The proposed PAPR reduction method is based on modification of the OFDM system in conjunction with the selective interleaving and the compound method. The performance of the proposed method is tested with image data transmission over two types of channels. The modification in the OFDM is done using double RS channel coder with optimum UEP to achieve higher quality of the received image with optimum bit rate. The simulation results indicates that, the proposed method reduce the PAPR/frame more than the conventional systems. Moreover, the overall distortion is less than the traditional OFDM system. The proposed optimum UEP improves the PSNR of the received image by 13.36 dB over the EEP coding method. The OFDM structure with selective interleaving improved the PSNR value of the received image by 10.2693 dB than the OFDM structure with compound method. Athough the selective interleaving method needs side information due to the optimization prosseing the transmission bit rate is less than that of the original OFDM bit-rate.

**Usama Sayed** received his B.Sc. and M.Sc. degrees from Assiut University, Assiut, Egypt in 1985 and 1993, respectively and Ph.D. degree from Czech Technical University in Prague, Czech Republic in 2000, all in electrical engineering. From 1988 to 1996, he was at Faculty of Engineering, Assiut University, working as an Assistant Lecturer. From February 1997 to July 2000, he was research assistant in the department of Telecommunications Technology at the Czech Technical University in Prague, Czech Republic. From December 1999 to March 2000, he was research assistant in the University of California In Santa Barbara (UCSB), USA. From November 2001 to April 2002, he was a post Doctoral Fellow with the Faculty of Engineering, Czech Technical University in Prague, Czech Republic. Since February 2006, he has been an Associate Professor with the Faculty of Engineering, Assiut University, Egypt. He authored and co-authored more than 75 scientific papers. Usama has been selected for the inclusion in 2010 Edition of the Marquis Who's Who in the World. His research interests include telecommunication technology, wireless technology, wireless Networks, RFID, image coding, speech coding, statistical signal processing, blind signal separation, and video coding.

**Hamada Ahmed Hamada** Demonstrator Communications & Electronics, Department Faculty of Engineering Aswan, South Valley University**,** research topic**,** wireless communication and DSP for modern communication system